\documentclass[11pt,twoside]{article}

\usepackage{asp2004}
\usepackage{epsf}
\usepackage{psfig}
\usepackage{lscape}
\usepackage{graphicx}
\usepackage{amssymb}

\newcommand{\dr}{\mbox{${\mbox d}$}}
\newcommand{\Nc}{\mbox{${N_{\rm cl}}$}}
\newcommand{\Mc}{\mbox{${M_{\rm cl}}$}}
\newcommand{\tdis}{\mbox{$t_{\rm dis}$}}

\newcommand{\tmax}{\mbox{$t_{\rm max}$}}
\newcommand{\Mdot}{\mbox{$\dot{M}$}}
\newcommand{\sigman}{\mbox{$\Sigma_{\rm n}$}}
\newcommand{\rhon}{\mbox{$\rho_{\rm n}$}}
\newcommand{\msun}{\mbox{${\rm M}_{\odot}$}}
\newcommand{\rh}{\mbox{$r_{\rm h}$}}
\newcommand{\rhoc}{\mbox{$\rho_{\rm cl}$}}

\begin{document}
\title{Cluster Disruption: Combining Theory and Observations}   
\author{Nate Bastian$^1$ and Mark Gieles$^2$}   
\affil{$^1$Department of Physics and Astronomy, University College London,
              Gower Street, London, WC1E 6BT, United Kingdom \\
       $^2$Astronomical Institute, Utrecht University, 
              Princetonplein 5, NL-3584 CC Utrecht, The Netherlands}    

\begin{abstract} 
We review the theory and observations of star cluster disruption.
The three main phases and corresponding typical timescales of
cluster disruption are: {\it I) Infant Mortality} ($\sim10^7$~yr),
{\it II) Stellar Evolution} ($\sim10^8$~yr) and {\it III) Tidal
relaxation} ($\sim10^9$~yr). During all three phases there
are additional tidal external perturbations from the host galaxy.  In
this review we focus on the physics and observations of Phase I and on
population studies of Phases II \& III and external perturbations
(concentrating on cluster-GMC interactions).  Particular attention is
given to the successes and short-comings of the Lamers cluster
disruption law,  which has recently been shown to stand on
a firm physical footing.

\end{abstract}



\section{Introduction}

The vast majority (perhaps all) of stars are formed in a clustered
fashion. However, only a very small percentage of older stars are
found in bound clusters.  These two observations highlight the
importance of clusters in the
star-formation process and the significance of cluster disruption.
The process of cluster disruption begins soon after, or concurrent
  with, cluster formation.
\cite{lada03} found that $\lesssim10\%$ of stars formed in embedded clusters
end up in bound clusters after $\sim10^{8}$~yr.
\cite{whitmore03} and \cite{fall05} have shown that at least 20\%, but perhaps
all, star formation in the merging Antennae galaxies is taking place
in clusters, the majority of which are likely to become unbound.  The case
is similar in M51, with $>60\%$ of all young ($<10$~Myr) clusters
likely to be destroyed within the first 10s of Myr of their lives
\citep{bastian05}.  On longer timescales, \cite{oort58} and
\cite{wielen71} noted a clear lack of older ($>$~few Gyr) open
clusters in the solar neighbourhood and \cite{bl03} found a strong absence
of older clusters in M51, M33, SMC, and the solar neighbourhood.

The lack of old open clusters in the solar neighbourhood is even more
striking when compared with the LMC, which contains a significant
number of `blue globular clusters' with ages well in excess of a Gyr
(e.g. \citealt{1966MNRAS.134...59G, degrijs06}).  This difference can
be understood either as a difference in the formation history of
clusters or as a difference in the disruption timescales.  This
later scenario was suggested by \citet{hodge87}, who
directly compared the age distribution of Galactic open clusters and
the SMC cluster population. He noted that there are $10-15$ times more
clusters with an age of 1 Gyr in the SMC as compared to the solar
neighbourhood (when normalising both populations to an age of
$10^8$~yr) and concluded that disruption mechanisms must be less
efficient in the SMC.

Much theoretical work has gone into the later scenario, with both
analytic and numerical models of cluster evolution predicting a strong
influence of the galactic tidal field on the dissolution of star
clusters (for a recent review see \citealt{baumgardt06}).  Only
recently has there been a large push to understand cluster disruption
from an observational standpoint in various external potentials,
making explicit comparison with models
\citep{bl03,lamers05a,lamers05b,gieles05a,lamers06}.

We direct the reader to the review by Larsen in these proceedings for
a historical look at the observations and theory of cluster
disruption.

\subsection{Phases of cluster disruption}
\label{subsec:phases}

\begin{figure}
\begin{center}
\includegraphics[height=8cm]{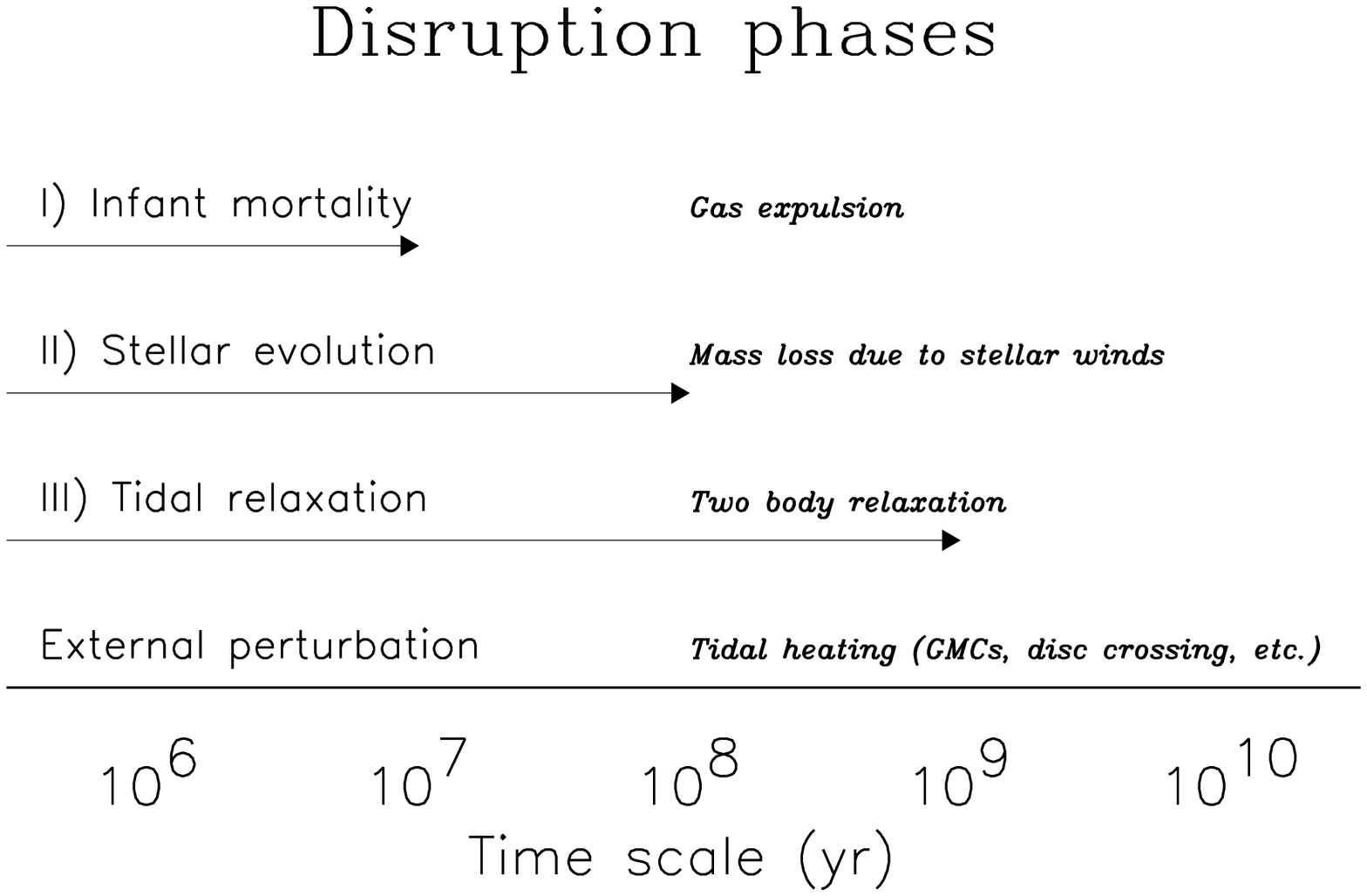}
\caption{Schematic overview of the three phases of cluster disruption
considered and the responsible physics that drives the disruption. }
\label{fig0}       
\end{center}
\end{figure}

While cluster disruption is a gradual process with several
different disruptive agents at work simultaneously, one can
distinguish three general phases of cluster mass loss and disruption.
As we will see, a large fraction of clusters gets destroyed during the
{\it primary}
phase. The main phases and corresponding typical timescales of
cluster disruption are: {\it I) Infant Mortality} ($\sim10^7$~yr),
{\it II) Stellar Evolution} ($\sim10^8$~yr) and {\it III) Tidal
relaxation} ($\sim10^9$~yr). During all three phases there
are additional tidal external perturbations from e.g. giant molecular
clouds (GMCs), the galactic disc and spiral arms that heat the cluster
and speed up the process of disruption. However, these external
  perturbations operate on longer timescales for cluster populations
  and so are most important in Phase III. In Fig.~\ref{fig0} we
schematically illustrate the three Phases of disruption and the
involved time-scales. Note that the number of disruptive agents
decreases in time.

 In this review we will focus on the physics and observations of Phase
I as well as on recent population studies aimed at understanding
Phases II and III on a statistical basis.  For a recent review on the
physics of Phases II and III, we refer the reader to
\citet{baumgardt06}.

Before proceeding, it is worthwhile to consider our definition of a
cluster. \cite{schweizer06} defines a cluster to be a {\it
gravitationally bound} stellar association which will survive for
10--20 crossing times.  This definition implies that the stars provide
enough gravitational potential to bind the cluster and ignores the
role of gas in the early evolution of clusters.  In this review, we
will define a cluster as a collection of gas and stars which was {\it
initially gravitationally bound}.  The reason for this definition will
become evident in Section~\ref{infantmortality}

\section{Infant Mortality}
\label{infantmortality}

Recent studies on the populations of young star clusters in M51
\citep{bastian05} and the Antennae galaxies \citep{whitmore03,fall05}
have shown a large excess of star clusters with ages less than $\sim$10~Myr
with respect to what would be expected assuming a constant cluster
formation rate.  The fact that open clusters in the solar
neighbourhood display a similar trend \citep{lada03} has led to the
conclusion that this is a physical effect and not simply that we are
observing these galaxies at a special time in their star-formation
history.  If one adopts this view, then we are forced to conclude that
the majority (between 60-90\%) of star clusters become unbound when
the remaining gas  (i.e. gas that is left-over from the star formation
process) is expelled.  These clusters will survive less than a few
crossing times.

\subsection{Gas expulsion}

Suppose that a star cluster is formed out of a sphere of gas with an
efficiency $\epsilon$, where $\epsilon = M_{stars}/(M_{stars} +
M_{gas})$.  Further suppose that the gas and stars are initially in
virial equilibrium.  If we define the virial parameter as
$Q=-2T/W$, with $T$ the kinetic energy and $W$ the potential energy,
virial equilibrium implies $Q=1$. Finally, suppose that the
remaining gas is removed on a timescale faster than the crossing time
of stars in the cluster.

In such a scenario the cluster is left in a super-virial state
after the gas removal, with $Q=1/\epsilon$, and the star cluster will
expand since the binding energy is too low for the stellar
velocities. The expanding cluster will reach virial equilibrium after a
few crossing times, but only after a (possibly large) fraction of the stars
have escaped.  
This process has been shown to remove a significant amount of the
stellar mass of a cluster, and if $\epsilon < 0.3$ the entire cluster
will become unbound on a timescale of 10s of Myr
\citep{tutukov78,goodwin97a,goodwin97b,kroupa02,boily03a,boily03b,bg06}.

\begin{figure}[!ht]
\hspace{-0.5cm}
\includegraphics[height=9cm]{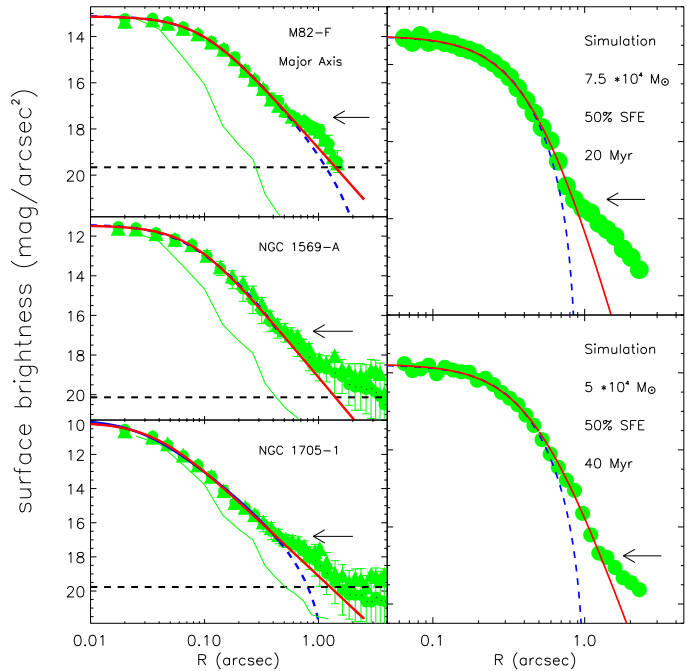}
\caption{Surface brightness profiles for
  three young clusters (left - M82-F, NGC~1569-A, and NGC~1705-1) and
  two $N$-body simulations which include the rapid removal of gas which
  was left over from a non-100\% star-formation efficiency (right).
  The solid (red) and dashed (blue) lines are the best fitting
  EFF~\citep{eff} and King~\citep{king} profiles respectively.  Note the
  excess of light at large radii with respect to the best fitting EFF
  profile in both the observations and models.  This excess light is
  due to an unbound expanding halo of stars caused by the rapid
  ejection of the remaining gas after the cluster forms.  {\it Hence,
    excess light at large radii strongly implies that these clusters
    are not in dynamical equilibrium.} For details of the modeling
  and observations see \cite{bg06} and \cite{gb06}.}
\label{fig1}       
\end{figure}

Rapid gas removal of the type discussed above leaves distinct
observables. In Figure~\ref{fig1} we show the surface brightness 
profiles of three young clusters (left panels) as well as two results
of $N$-body simulations (right panels) of clusters including the
effects of rapid gas removal.  All three young clusters show an excess
of light at large radii with respect to the best fitting
EFF~\citep{eff} or \citet{king} profiles.  This is in good agreement
with the predictions of the simulations, in which an unbound halo of
stars is removed (although still appearing to be associated with the
cluster for 10s of Myr) due to the rapid change of the gravitational
potential \citep{bg06}.  Such excess light at large radii has also
been found in young clusters in the Antennae galaxies
\citep{whitmore99}.  \citet{gb06} show that for values of $\epsilon$
of 0.1 and 0.6, clusters will lose 75\% and 10\% of the stellar mass
respectively within the first $\sim20$~Myr of their lives.

Thus we see that this is an extremely efficient way to rapidly
disperse stars from young clusters into the field.  This mechanism
provides a natural explanation for the observed diffuse UV light
in the field of starburst galaxies \citep{tremonti01,chandar05} and
supports the scenario of these authors that this light is due to
rapidly dispersing young clusters.

Whether or not a cluster survives this phase, and hence more than 10--20
crossing times, is largely dependent on the star-formation efficiency
of the GMC core in which the cluster formed.  Thus, two clusters with exactly
the same parameters (radius, mass, metallicity, external potential
field, etc) may experience two radically different evolutionary paths
if their star-formation efficiencies are different.  \cite{gb06} have
used the internal dynamical properties of young clusters in order to
estimate their $\epsilon$-values.  No clear trend of $\epsilon$ on cluster
(stellar) mass or radius was found.

\subsection{Interpretation of the properties of young clusters}
Even if a cluster survives the gas removal phase, this phase can
significantly effect the observed properties of the cluster.  Hence,
deducing the initial properties of a cluster from its current state is
not trivial. \cite{kroupa02} have noted the strong effect of residual
gas removal on inferring the initial stellar mass of a cluster, while
\cite{gb06} have refined the mass loss estimates and shown that
measurements of the current radii of young clusters may not reflect
the initial nor the final value. Additionally, \cite{gb06} show that
this effect can mimic stellar IMF variations in young clusters.

\section{Population Studies of Cluster Disruption}
\label{sec:populations}

The clusters that have survived the gas removal phase are subject to
disruption Phases II and 
III (\S~\ref{subsec:phases}) as well as tidal effects. Disruption due
to these effects can be studied on individual clusters, of
which the recent observations of the dissolving globular clusters Palomar 5
\citep{2001ApJ...548L.165O} are probably the most spectacular
example. However, much can be learned by approaching this problem
from a cluster population point of view.

\subsection{The Lamers disruption law}

Suppose that clusters are formed continuously with a constant cluster
formation rate (a constraint which we can relax later).  Also, we will
assume that we know the cluster initial mass function (here taken to
be a power law of the form $\Nc\dr \Mc \propto \Mc^{-\alpha}\dr \Mc$
with ${\alpha} = 2$
(e.g.~\citealt{1999ApJ...527L..81Z, degrijs03}) and that
clusters can be detected down to a known magnitude limit.  Finally, we
will assume that the disruption time of a cluster depends on the
cluster mass, such that more massive clusters survive longer
(on average) than lower mass clusters.  For this final assumption we
will adopt a function of the form:

\begin{equation}
\tdis = t_{4}(\Mc/10^{4}M_{\odot})^{\gamma},
\label{eq:Ihavebeentypingthisequationtoomuchinmylife}
\end{equation}
 where $t_{4}$ is the disruption
time of a $10^4\,M_{\odot}$ cluster and $0<\gamma<1$ \citep{bl03}.
The beauty of this formulation is that it only has two variables,
$t_{4}$ and $\gamma$, and as we will see, provides extremely good fits
to observations.

\subsection{Application to various cluster populations}
\label{subsec:appl_pops}

The formulation provided above, when combined with the given
assumptions, allows for the parameters $t_{4}$ and $\gamma$ to be
found from age and mass distributions of clusters.  The first survey
using this technique was carried out by \cite{bl03} on cluster
populations in M51, M33, the SMC, and the solar neighbourhood.
They made a {\it sudden disruption} assumption, meaning that the
cluster is in the sample with its initial mass until \tdis, when it is
disrupted. The somewhat surprising result from this study was that,
while $\gamma$ had more or less the same value in all environments
studied ($\langle\gamma\rangle = 0.62 \pm 0.06$), $t_{4}$ varied by over two orders
of magnitude, with values of $\sim80$~Myr in the central regions
of M51 to $\sim8$~Gyr in the SMC.

\begin{figure}
\begin{center}
\includegraphics[width=6.75cm]{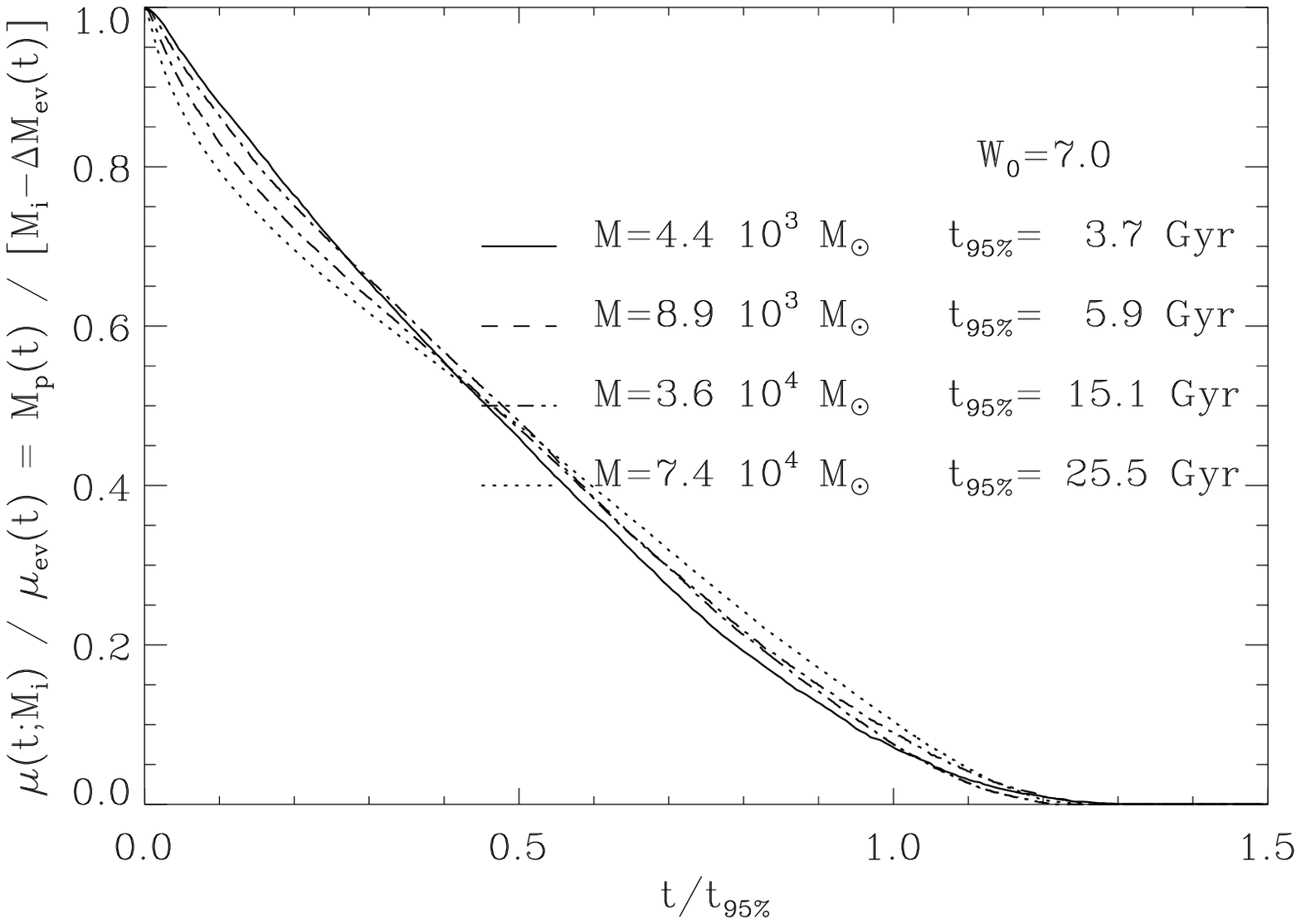}\includegraphics[width=6.75cm]{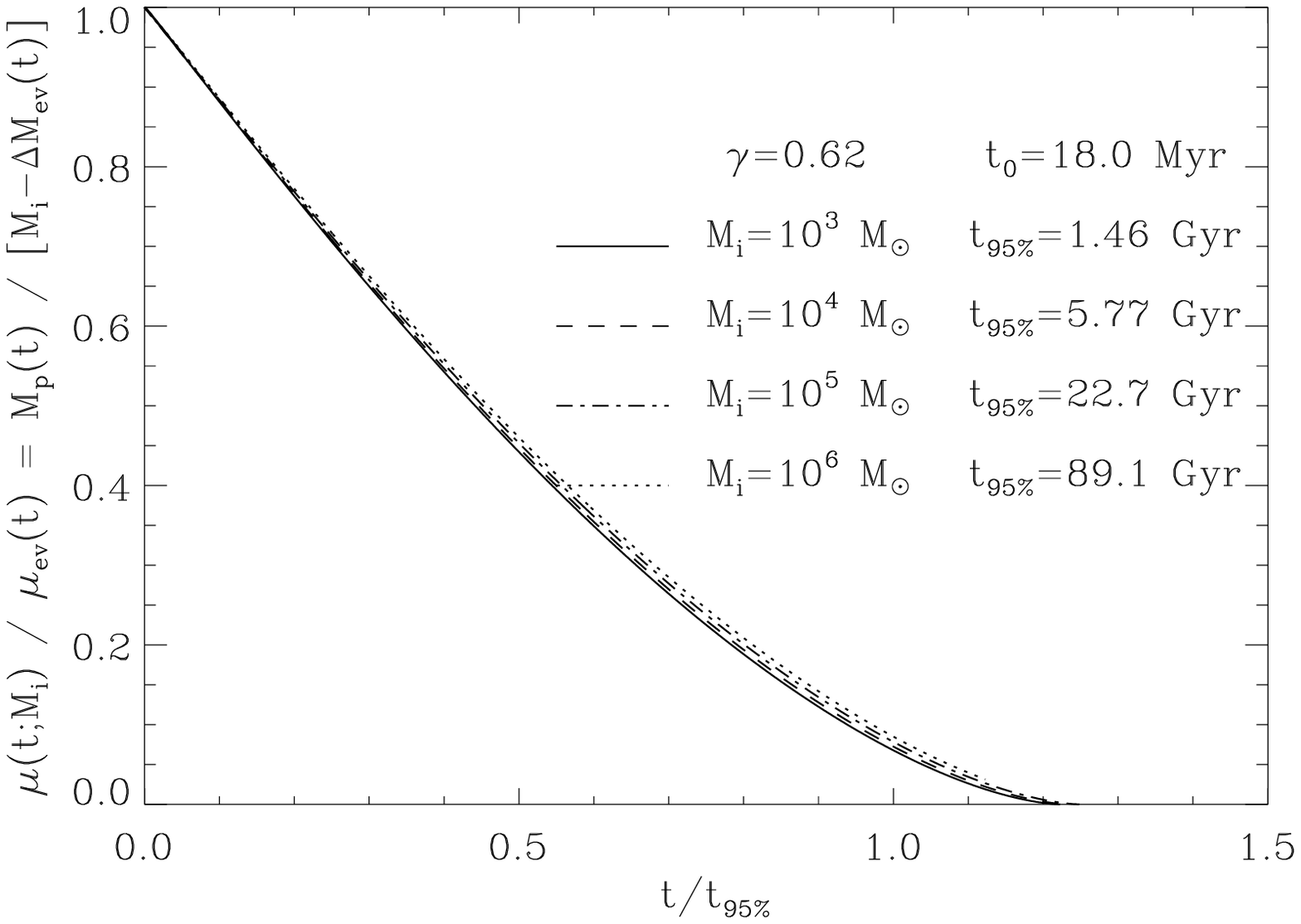}
\caption{Comparison between the mass loss following from the $N$-body
simulations of clusters with different number of stars, different
concentration and on different orbits (left). The mass loss due to
stellar evolution is not shown. In the right panel the analytical
model of \citet{lamers05a} is shown.}
\label{dm_lamers}       
\end{center}
\end{figure}

The simple {\it sudden disruption} assumption was improved in a
more recent model by \citet{lamers05a}, where a gradual loss of
cluster mass was implemented. They assumed that the cluster mass
decreases exponentially with a time-scale that decreases as the
cluster mass decreases. This is done by saying that the mass loss per
unit time (\Mdot) relates to \tdis\ as:

\begin{equation}
\Mdot=\Mc/\tdis\propto\Mc^{1-\gamma},
\label{eq:dm} 
\end{equation}
with $\tdis$ from
Eq.~\ref{eq:Ihavebeentypingthisequationtoomuchinmylife}.  This
very simple analytical description for cluster mass loss shows
remarkably good agreement when compared to the mass loss following
from the detailed $N$-body simulations of \citet{bm03}. In
Fig.~\ref{dm_lamers} we show a direct comparison of $\Mdot$ from the
$N$-body simulations of clusters with different density profiles and
on different orbits (left) and the above mentioned analytical model of
\citet{lamers05a} (right). 

In both graphs the time is normalised to \tdis\ and only the mass loss due
to stars escaping the cluster is shown, i.e. mass loss due to
stellar evolution (SEV) is not shown. In addition, there is a coupling between the two types of mass loss: if
stars loose mass, the cluster will expand and more stars are pushed over the
tidal boundary. The simulations of \citet{bm03} considered SEV, therefore,
their \Mdot\ results shown in Fig.~\ref{dm_lamers} do include tidal \Mdot\
induced by SEV. For this reason we can simply add the mass loss due to SEV,
taken from an SSP model, to Eq.~\ref{eq:dm}.


In a series of follow-up works, it has been shown that the similarity
of the value of $\gamma$ in various environments strongly implies a
uniformity in the cluster disruption process, while the varying values
of $t_{4}$ is due to the different tidal field strengths {\it (and gas
contents)} of the galaxies studied. Galaxies with strong tidal fields,
as, for example, derived from their rotation curves, having shorter disruption times
\citep{lamers05b}.  Comparison with results of realistic $N$-body
models performed by \citet{bm03} have placed this empirical disruption
law on a solid physical footing \citep{lamers05a}. \citet{lamers05a} have also derived a formula for the
predicted mass and age distributions of cluster samples
that includes both stellar evolution and disruption
for any star formation history.

\begin{figure}[!ht]
\begin{center}
\includegraphics[width=6cm]{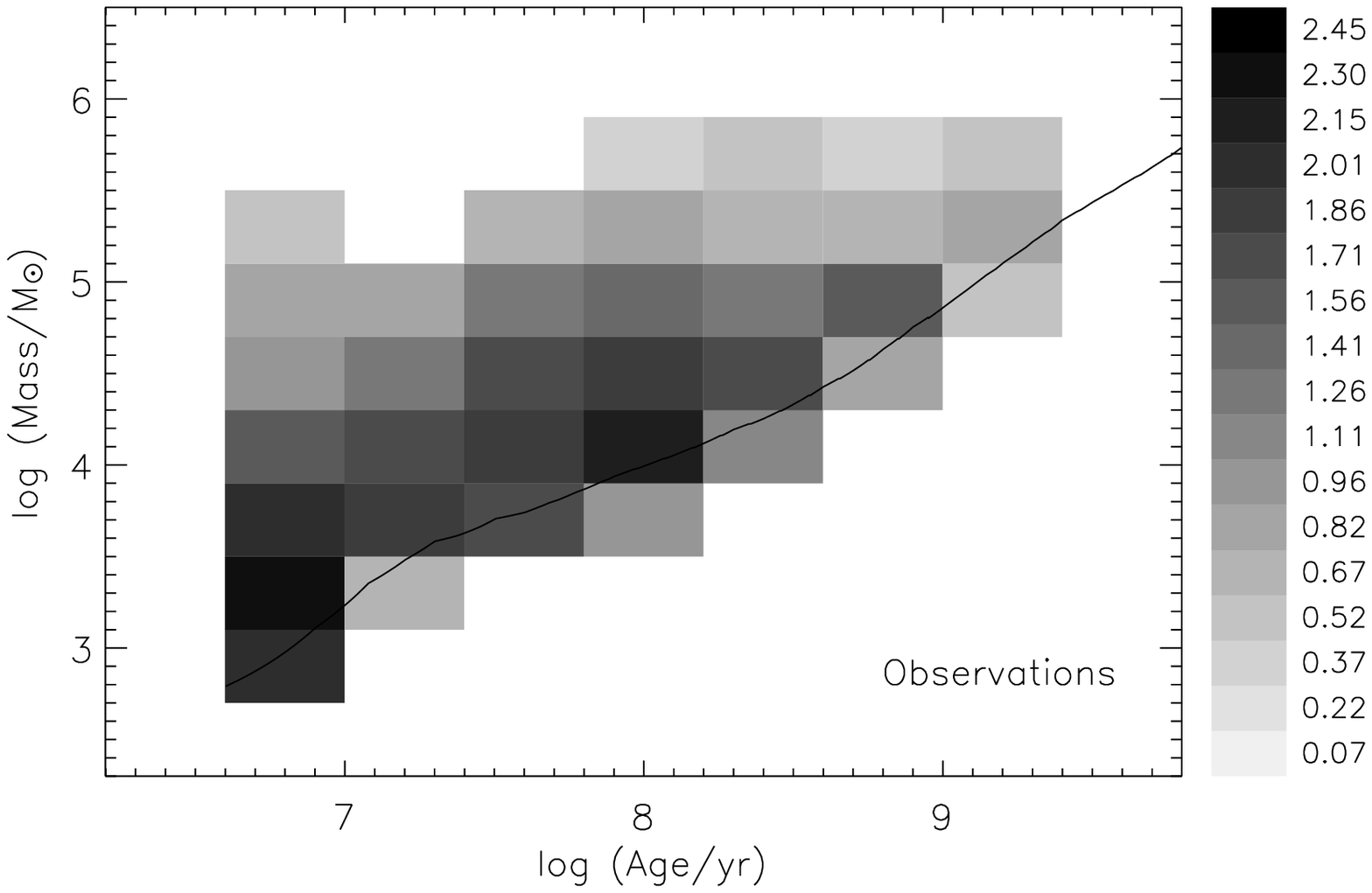}\includegraphics[width=6cm]{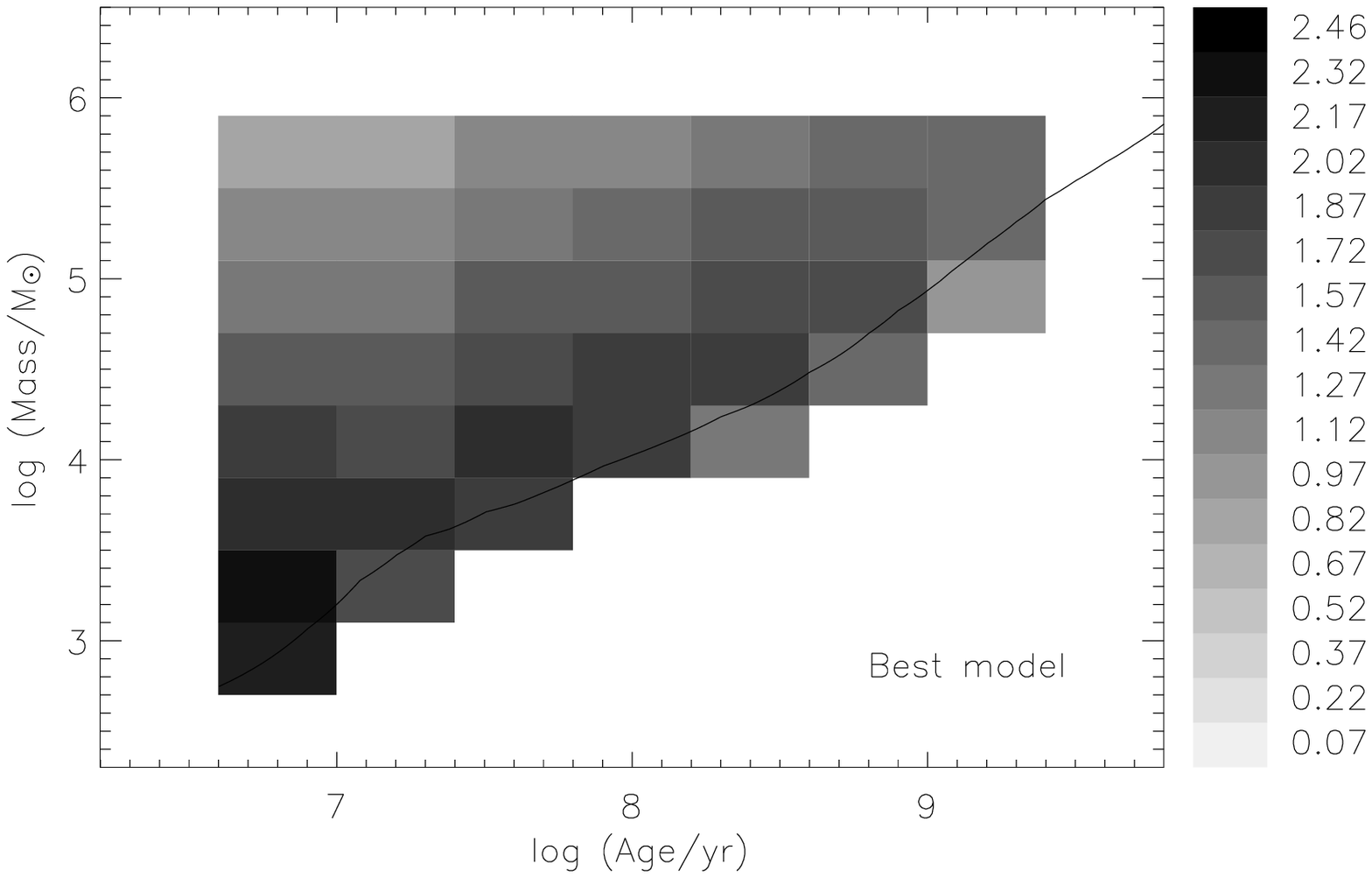}
\includegraphics[width=6cm]{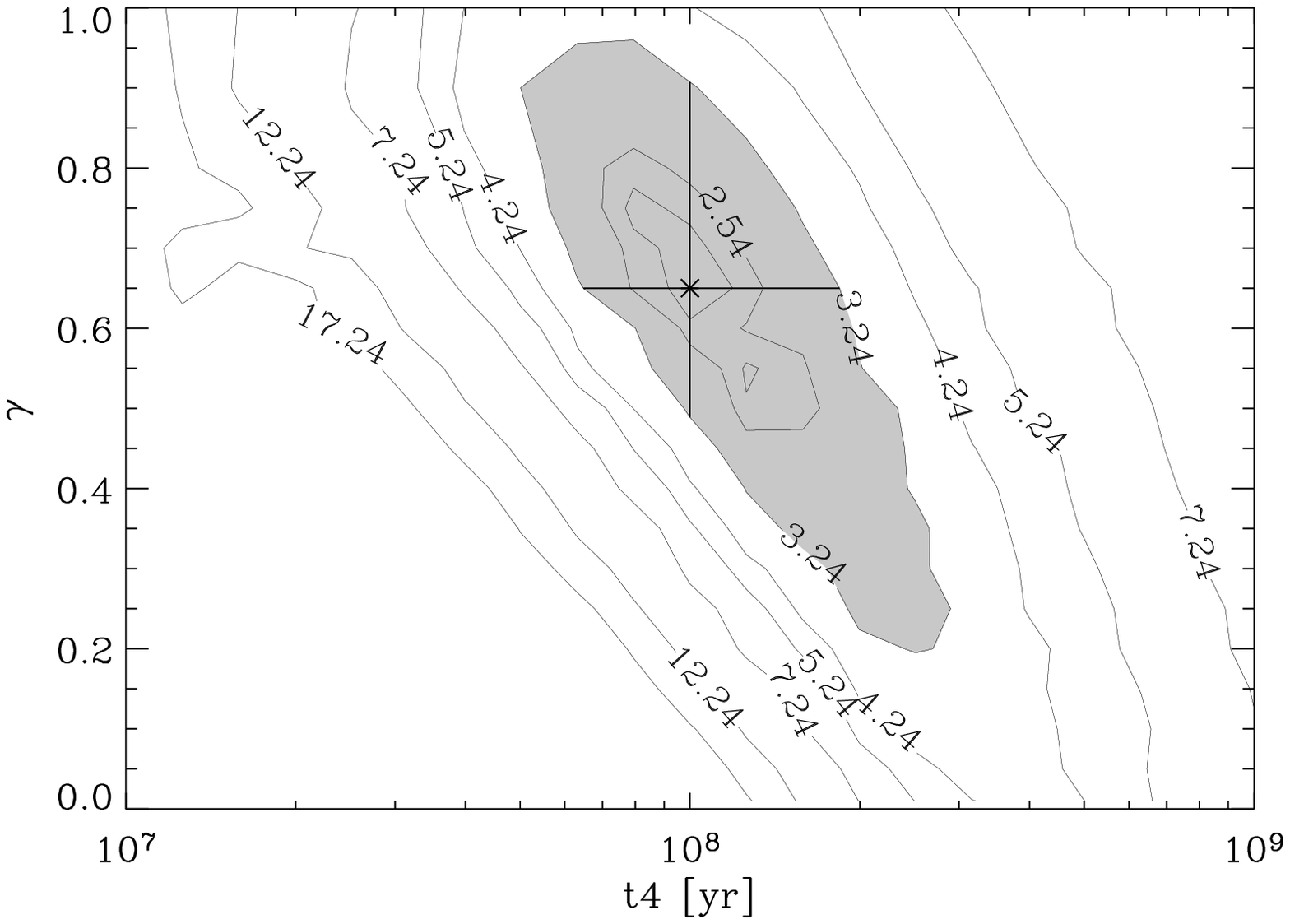}
\caption{{\bf Top left panel:} The observed age-mass diagram of clusters
  in M51.  The shading represents the logarithm of the number of
  clusters found within the corresponding box in age-mass space.  The
  solid line represents the detection limit of the sample. {\bf Top
  right panel:} The best fitting cluster population synthesis model
  which includes the detection limit of the observations and has fit
  on $\gamma$ and $t_{4}$.  {\bf Bottom:}  The $\chi^2_{\nu}$
  diagram in $\gamma$-$t_{4}$ space.  The best fitting model is marked
  with an 'x', while the accepted fits ($\chi^2_{\nu} <
  \chi^2_{\nu,{\rm best}} + 1$) are shaded.  Reproduced
  from \citet{gieles05a}.}
\label{fig2}       
\end{center}
\end{figure}

\cite{gieles05a} inserted the Lamers disruption law into a cluster
population synthesis model. This method has two distinct advantages
over the earlier formulations. The first is that it removes the
requirement of a constant cluster formation rate, and second, it uses
the age and mass distributions together to find $\gamma$ and $t_{4}$.
The case of M51 is shown in Fig.~\ref{fig2}.  One first begins by
constructing an observed number density grid in age-mass space
(upper-left panel where the shading corresponds to the logarithm of
the number of clusters found within that cell).  Then one generates a
large number of models with different values of $t_{4}$, $\gamma$,
(time dependent) cluster formation rates, etc. and compares these
models with the observed grid.  The resultant $\chi^2_{\nu}$ diagram
is shown in the bottom panel of Fig.~\ref{fig2}.  The best fit
model is shown in the top right panel of Fig.~\ref{fig2}.

This cluster population synthesis (CPS) technique, also used in a
similar way by \cite{dolphin02} to derive the properties of the
cluster population in the galaxy NGC~3627, holds great promise in
disentangling the 
myriad of effects present in cluster populations. In principle, the
dependences of cluster size, galactocentric radius, 
star-formation efficiency dependent infant mortality rates, or
alternative cluster disruption models can be taken into account by
this technique.  For this
technique to be fully exploited one needs large samples of cluster
populations with known ages and masses.  Datasets suitable for
these kind of studies are beginning to be collected and released. Several
face-on spiral galaxies have been imaged in multiple filters with the
high resolution/wide field {\it HST/ACS} camera (e.g
\citealt{2006AJ....132..883B} for M101 and
\citealt{2006A&A...446L...9G} for M51).  

\subsection{Possible objections}

It is worth noting possible objections to the Lamers disruption law.
The first comes from \cite{fall05} who find that in the Antennae
galaxies the number of clusters decreases in time ($\tau$) as $\dr
N/\dr \tau \propto \tau^{-1}$, independent of cluster mass.  This
may be explained if the disruption timescale $t_{4}$ due to tidal
field effects (e.g. Phase II \& III) is greater than or similar to
the maximum age in the sample. The cluster disruption due to tidal
effects would not yet be present in the \cite{fall05} sample, instead
the decrease in cluster numbers would be the result of infant
mortality and the fading of clusters.  Studies of infant mortality in
M51 also suggest that the effect is mass independent
\citep{bastian05}.  In fact, if infant mortality was not (mostly) independent
of cluster mass we would expect the embedded cluster mass function to be
significantly different from the optically selected cluster mass function.

\begin{figure}
\begin{center}
\includegraphics[height=4cm]{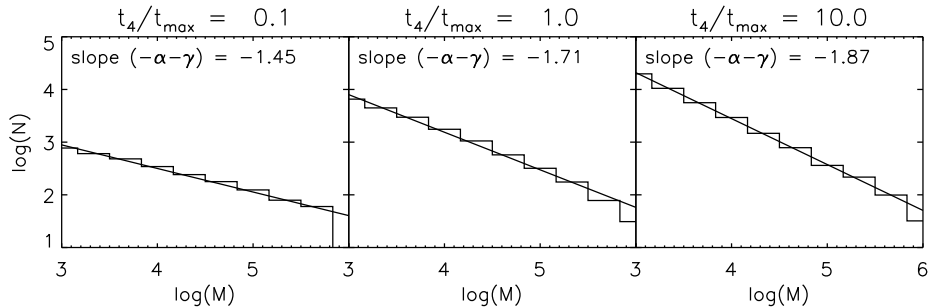}
\caption{Representation of how the slope of the mass function changes
  as a function of the ratio between the disruption time (\tdis) and
  the age of the oldest cluster in the sample (in this case assuming a
  constant cluster formation rate).  Note that if \tdis~approaches or exceeds 
  \tmax~the slope of the mass function approaches the initial mass
  function value (i.e.~$-2$).}
\label{fig:mf}       
\end{center}
\end{figure}

In Fig.~\ref{fig:mf} we show the dependence of the mass function slope
of a multiple age cluster population on the ratio of the $t_4$  and
the maximum age of the 
cluster in the sample (\tmax).  Clusters were created continuously
over 1 Gyr with an initial mass function of a power-law with index
$-2$.  The Lamers disruption law was applied in the same way as in
\citet{gieles05a}.  The important thing to take away from this figure
is that as \tdis~approaches and exceeds \tmax~the mass function is
less affected by disruption and so it retains its initial form,
i.e. the right panel in Fig.~\ref{fig:mf} probably applies to the
\citet{fall05} sample.

A second observation seemingly contradicting the Lamers disruption law
is that of \cite{chandar06} who find an intermediate age
($\sim4-7$~Gyr) globular cluster in M33.  In M33, \cite{lamers05b}
find a $t_{4}$ value of $\sim600$~Myr, implying a disruption time of
$\sim2.5$~Gyr for a $10^5~M_{\odot}$ cluster.  However, as the authors
note, the value of $t_{4}$ derived by \cite{lamers05b} was presumably
of the thin disk of the galaxy, and if the intermediate-age cluster is
part of the thick-disk or halo of the galaxy then the expected value
of $t_{4}$ would be significantly larger than that quoted.
Additionally, it should also be noted that the mass derived by
\citet{chandar06} is the {\it present} mass of the cluster.  The
cluster presumably started with a much higher mass and disruptive
effects have brought this cluster into its current state.  If the
present mass of the cluster is $1\times10^5$~\msun, then its initial
mass (after infant weight loss) would have been $5\times10^5$~\msun
(assuming an age of 5~Gyr), using the
value of $t_{4}$ for M33 given by \cite{lamers05b}.
 
\section{External Disruption Effects: GMCs and Spiral Arms}
\label{sec:external}

\subsection{The disruption time due to external perturbations}
\label{subsec:tdis_external}

\begin{figure}[!t]
\begin{center}
\vspace{-1cm}
\includegraphics[width=9cm]{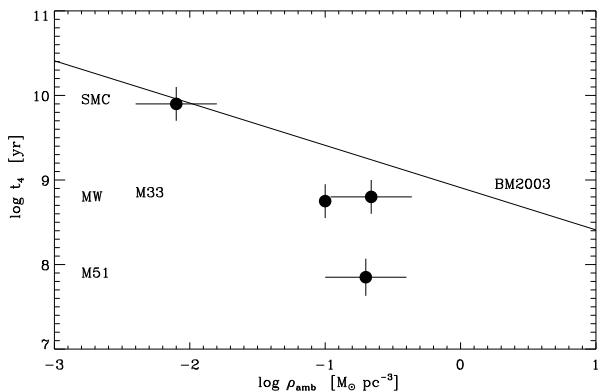}
\caption{Comparison between the observed (filled circles) and
  predicted (solid line; from the $N$-body models of \citealt{bm03}) disruption time of
  a $10^4~M_{\odot}$ cluster, $t_{4}$, as a function of the mean
  density $\rho_{\rm amb}$ in $M_{\odot}$pc$^{-3}$ of the host
  galaxy.  Reproduced from \cite{lamers05b}.}
\label{fig3}       
\end{center}
\end{figure}

As more and more galaxies (and environments) have their
characteristic disruption timescales measured, it is useful to compare
the results to $N$-body models in order to check for consistency between the
two.  This was done in \cite{lamers05b} who compared the $t_{4}$
values derived for the SMC, M33, M51 and the solar neighbourhood to
the $N$-body models of \cite{bm03} and \cite{pz98,pz02} which sample a
large range in the ambient densities of the host galaxies.  Their
results, shown in Fig.~\ref{fig3}, are intriguing. While the
predicted and observed disruption time of the SMC are in excellent
agreement, the disruption times of the Galaxy, M33 and M51 are observed
to be much shorter than predicted by $N$-body models.  This result is
particularly surprising given the fact that the mass loss predictions
of a single cluster are in excellent agreement between the Lamers
empirical description and that given by $N$-body models
(\citealt{lamers05a} and Fig.~\ref{dm_lamers}).

\begin{figure}
\begin{center}
\vspace{0cm}
\includegraphics[width=14cm]{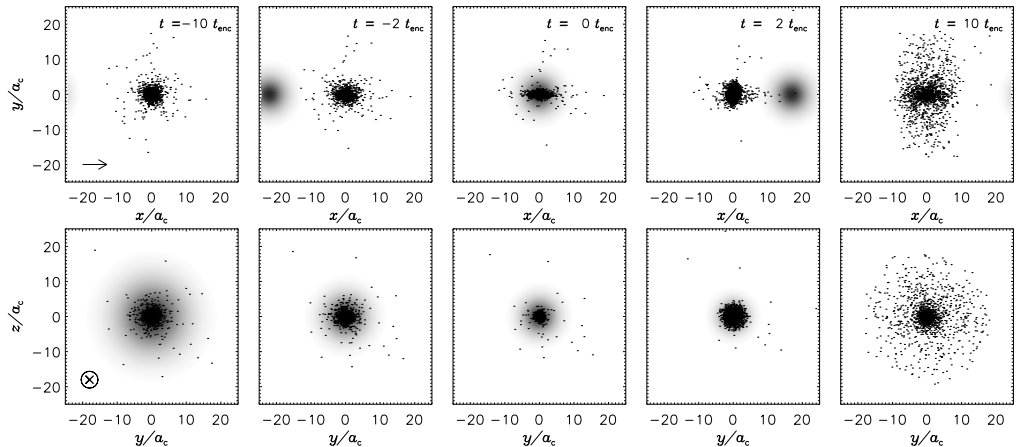}
\caption{Snapshots of a star cluster undergoing a head-on encounter
  with a GMC of M$_n$ = $10^4~M_{\odot}$ with $V_{\rm max} =
  20\sigma_{\rm rms}$. {\bf Top:} The motion of the GMC is along the
  x-axis and the line of sight is perpendicular. {\bf Bottom:}  The
  motion of the GMC is into the page and the line of sight is along
  the GMC trajectory. The arrows in the left-hand lower corner 
  of the left-hand panels are parallel to the direction of motion of
  the GMC. The GMC is shown with grey shades based on the surface
  density of a GMC with a$_n$ = 5.8a$_c$. The time with respect to the
  moment of encounter is indicated in each panel of the top row.  See
  \citet{gieles06c} for a description of the methods and parameters used.}
\label{fig4}       
\end{center}
\end{figure}

Thus, we are left to ask, what physical effects are not included in
the $N$-body models that may be responsible for disrupting clusters?
The $N$-body models used in the comparison were carried out in a
smooth logarithmic potential which does not realistically represent
the thin disk components of disk galaxies. \cite{gieles06a,gieles06c}
have attempted to add encounters with giant molecular clouds (GMCs)
and spiral arm passages to the $N$-body models. In Fig.~\ref{fig4} we
show an example of a cluster-GMC encounter (from
\citealt{gieles06c}). The parameters of this run are for typical open
clusters and GMCs in the solar neighbourhood. The top panels show an
edge-on view for five different time steps, while the bottom panels show
a view along the trajectory of the GMC. 

Encounters with GMCs present the most important external
perturbation which cause mass loss of star in clusters. \citet{gieles06c} find that \tdis\ due to encounters with
GMCs scales as

\begin{equation}
\tdis=2.0\,S\left(\frac{\Mc}{10^4\,\msun}\right)\left(\frac{3.75\,{\rm
pc}}{\rh}\right)^3{\mbox{Gyr}},
\label{eq:tdis_gmcs}
\end{equation}
where $S\equiv1$ for the solar neighbourhood and $S$ scales with the
surface density of individual GMCs ($\sigman$) and the global GMC
density ($\rhon$) as $S\propto(\sigman\rhon)^{-1}$. The scaling of $S$
with $\rhon$ implies that it does not matter if the molecular gas is
distributed over a large number of low mass clouds or a small number
of massive (giant) clouds. This makes it easy to estimate \tdis\ from
the observed molecular gas density. Indeed, for M51, where the
molecular gas density is about an order of magnitude higher than in
the solar neighbourhood, a \tdis\ from Eq.~\ref{eq:tdis_gmcs} of
150~Myr is predicted. This corresponds well with the value derived
from observations of $t_4=100-200$~Myr \citep{gieles05a}.

Note that Eq.~\ref{eq:tdis_gmcs} implies a scaling of \tdis\ with the
cluster density ($\rhoc$). This seems different than the dependence
with \Mc\ discussed before. However, there is only a very shallow
relation observed between cluster half-mass radius (\rh) and \Mc, of
the form $\rh\propto\Mc^{0.1}$ \citep{larsen04}. With this
relation, and Eq.~\ref{eq:tdis_gmcs}, it follows that for external
perturbations $\tdis\propto\Mc^{0.7}$, i.e. very close to the index of
$\gamma\simeq0.6$ found from observations discussed in
\S~\ref{subsec:appl_pops}. This suggests that the disruptive effect of
the tidal field and additional external perturbations can be added
linearly, resulting in a \tdis\ that depends on \Mc\ as
$\tdis\propto\Mc^{0.6}$. This can explain the large variation found in
the $t_4$ value derived from observations and the almost
constant $\gamma=0.6$ \citep{bl03}. In \S~\ref{sec:discussion} we
discuss some of the pitfalls of these results.

\subsection{Application to the open clusters in the Galaxy}
\begin{figure}[!ht]
\begin{center}
\includegraphics[width=6.75cm]{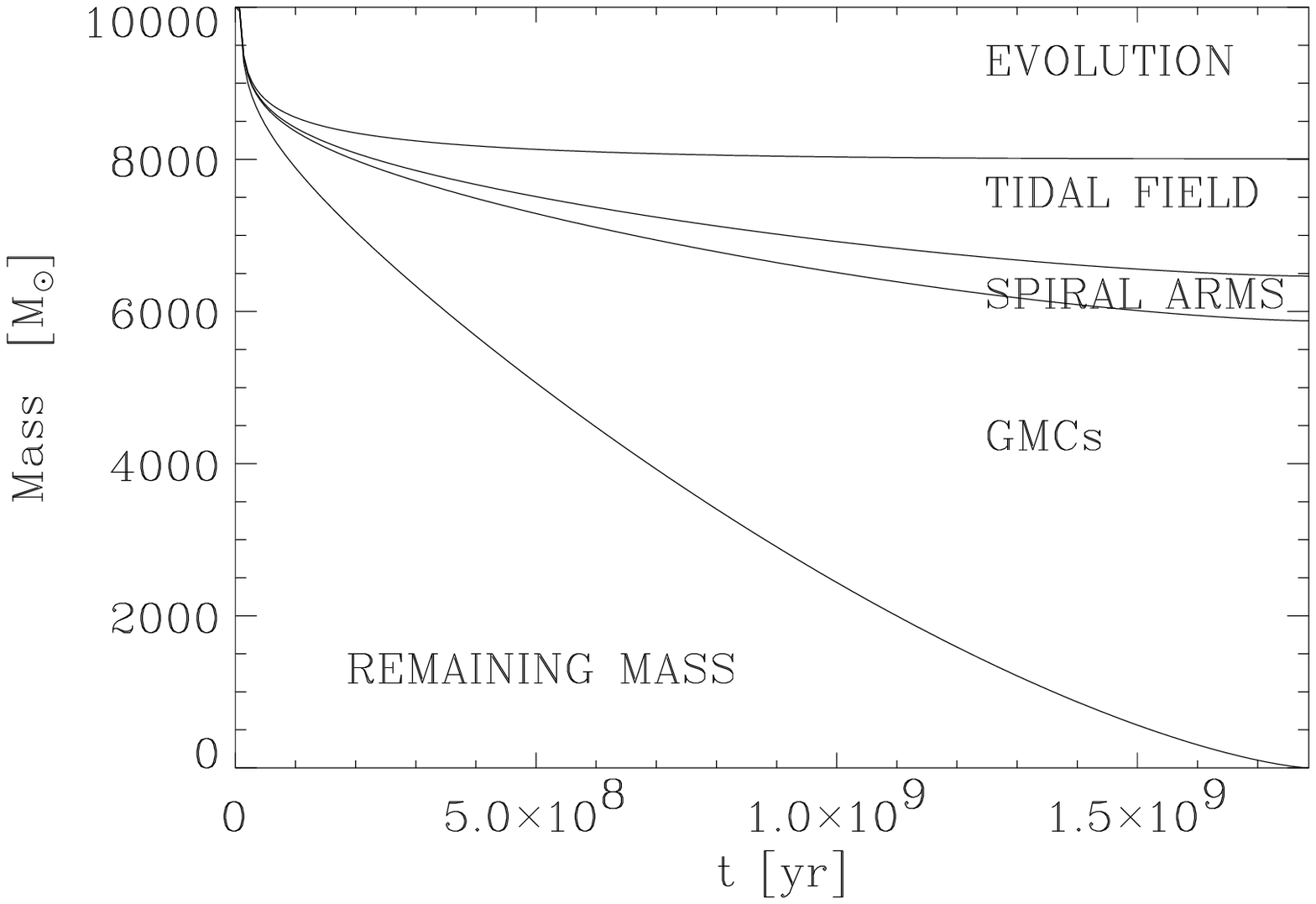}\includegraphics[width=6.75cm]{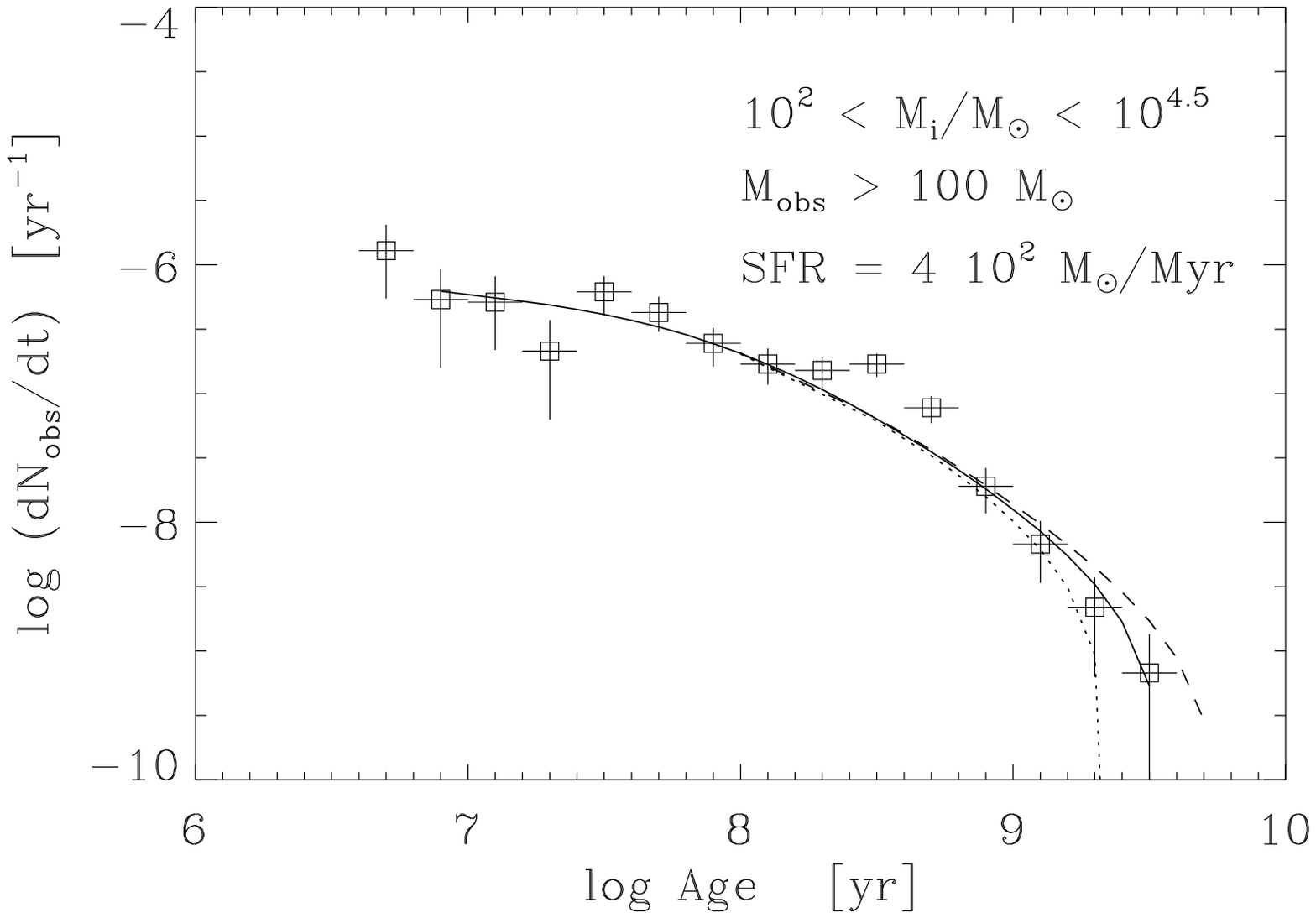}
\caption{{\bf Left:} The mass evolution of a $10^4\msun$ cluster due
to various disruptive effects. {\bf Right:} Comparison between the
observed age distribution of open clusters (from
\citealt{2005A&A...438.1163K}) and the predictions from
\citet{lamers06} for three different maximum masses. }
\label{fig5}       
\end{center}
\end{figure}

As seen in the proceeding sections, the observed disruption time of
star clusters in the solar neighbourhood is a factor of $\sim5$
shorter than predicted by $N$-body models.  The inclusion of spiral
arm passages and GMC encounters into $N$-body models is a promising
way to bring the predictions into agreement with the observations.
This was recently done by \cite{lamers06} who found excellent
agreement after the inclusion GMC encounters and spiral arm
passages. They assume that the different mass loss effects (stellar 
evolution, tidal field and external perturbations) can be added
linearly. Using the mass-radius relation of
\S~\ref{subsec:tdis_external} and the results from \citet{gieles06a}
and \citet{gieles06c} they analytically model the mass loss due to different
effects analytically. This is illustrated in the left panel of Fig.~\ref{fig5} (from \citealt{lamers06}). Based on this mass loss
description, the age distribution of open clusters in the solar
neighbourhood can be predicted (instead of fitted, as was done
hitherto). The results are shown in the right panel of Fig.~\ref{fig5}.

\section{Discussion}
\label{sec:discussion}

We showed in \S\S~\ref{sec:populations}\&\ref{sec:external} that
the simple Lamers disruption law can successfully explain the age and
mass distribution of young star clusters populations. Here we will
discuss other observations lending support to the Lamers law and some
of the standing problems and uncertainties of this scenario which
need further attention. 

\subsection{Independent checks on the disruption law}

It is reassuring to see that different datasets of various cluster
populations all come to similar conclusions regarding the disruption
of clusters.

\citet{degrijs06} use a variety of studies to look at the
cluster population of the LMC.  They also find a lack of old clusters
(with respect to what would be expected from a continuous cluster
formation rate) and derive $\gamma=0.56$, again in agreement with
other galaxies studied by \citet{bl03} and \citet{lamers05a}. Note
that a lower value of $\gamma$ is expected to be observed when the typical 
\tdis\ is of the same order as the oldest clusters in the sample
(Fig.~\ref{fig:mf}), as is the case in LMC. 

Outside the local group, the strongly interacting galaxy NGC~6745 has
been studied by \citet{degrijs03} who found evidence for mass
dependent disruption, with $\gamma=0.54$.  The rich cluster system of
the intermediate-age merger remnant NGC~1316 shows a clear bimodal
colour distribution, with the red component presumably being formed
during the merger.  \citet{goudfrooij04} showed, using deep {\it
  HST-ACS} images that if one breaks the red component into `inner'
and `outer' regions (with respect to the galactic centre), that the
outer region is a continuous power-law while the inner region shows a
power-law behavior at the high luminosity end and a flattening at the
low luminosity end.  The authors interpret this as evidence for
mass-dependent cluster disruption, although no attempt was made to
find the characteristic disruption timescale or the value of $\gamma$.

One standing problem with the Lamers disruption law, also present in
other studies on 
disruption, is whether or not an initial power-law cluster initial
mass function (CIMF) can be transformed into a log-normal distribution,
which is observed for old globular cluster populations.  The Lamers law can
create such a turnover, however the precise value of the turnover mass
should be dependent on the ambient density \citep{lamers05b}, meaning
that cluster disruption should be more efficient in the inner regions
of a galaxy than in the outer regions.  Thus, without fine tuning the
models (e.g. having the same disruption time at all radii due to large
radially dependent velocity anisotropies) one would expect a radially dependent
turnover peak in the globular cluster mass function, which is not
observed.  For a more detailed description of this problem, see the review by
Larsen in these proceedings.  Additionally, as noted by
\citet{waters06} the Lamers disruption over-predicts the number of
low-mass clusters when applied to old globular cluster populations.

\subsection{Caveats in the theoretical underpinning}
In \S~\ref{subsec:tdis_external} we showed that the scaling of \tdis\
with \Mc\ is a power-law with exponent $\gamma\simeq0.6$.  This scaling is
similar for two-body evaporation in a tidal field with external
perturbations, such as shocks by GMCs and spiral arms, and agrees well
with the observations. However, there are still some caveats in the
theory explaining this, mostly coming from questions regarding the
initial conditions of the simulations.

\begin{enumerate}

\item The first caveat stems from the relation between initial mass
and radius of the clusters used in the simulations.  If we
parameterize this relation as \rh\ $\propto$ \Mc$^\lambda$, then
\citet{bm03} use $\lambda = 1/3$, implying that their clusters fill
their tidal radius.  However, observations imply that $\lambda =
0-0.1$ (with a large scatter) \citep{larsen04,bastian05}, implying
that \rh\ is mostly independent of mass. This shallow relation implies
that massive clusters are not filling their tidal radius, which would
change the dependence of \tdis\ with \Mc\ \citep{tanikawa05}. 

\item 
In the derivation of $\gamma$ for external shocks
(\S~\ref{subsec:tdis_external}), only clusters in isolation were
considered. How would the presence of a tidal field affect this result?

\item  How does the relation for \tdis\ change if there exists a
  relation between the concentration parameter and mass of a cluster
  (i.e. as seen in the Galactic GCs reported by Larsen in these proceedings)?

\item Could an initial mass-radius relation with $\lambda=1/3$ be
  erased during the gas removal phase?

\item What is the effect of the initial mass/luminosity profile used
  in the simulations and how does it evolve?  e.g. are clusters born
  with EFF profiles which are converted into King profiles?  Does the
  cluster concentration alter its mass loss evolution?

\item How do the external perturbations and the galactic
tidal field cooperate? Can the mass loss due to both effects simply be
added linearly?

\end{enumerate}

\acknowledgements Both NB and MG are grateful to Henny for all of his
guidance, ideas and help.


\end{document}